\documentclass[12pt,a4paper]{article}

\usepackage{essv}
\usepackage{multicol}
\usepackage{multirow}
\usepackage{float}
\usepackage{tabularx}
\newcolumntype{Y}{>{\centering\arraybackslash}X}

\title{Scalable engine and the performance of different LLM models in a SLURM based HPC architecture}
\author{Anderson de Lima Luiz, Shubham Vijay Kurlekar, Munir Georges}
\affil{AImotion Bavaria, Germany \\
Technische Hochschule Ingolstadt\\
Esplanade 10, 85049, Ingolstadt, Bayern, Germany }
\email{\{anderson.delimaluiz, shubhamvijay.kurlekar, munir.georges\}@thi.de}

\begin{document}


\maketitle

\begin{abstract}
This work elaborates on a High performance computing (HPC) architecture based on Simple Linux Utility for Resource Management (SLURM) \cite{slurm} for deploying heterogeneous Large Language Models (LLMs) into a scalable inference engine. Dynamic resource scheduling and seamless integration of containerized microservices have been leveraged herein to manage CPU, GPU, and memory allocations efficiently in multi-node clusters. Extensive experiments, using Llama 3.2 (1B and 3B parameters) \cite{LLAMA32} and Llama 3.1 (8B and 70B) \cite{LLAMA31}, probe throughput, latency, and concurrency and show that small models can handle up to 128 concurrent requests at sub-50 ms latency, while for larger models, saturation happens with as few as two concurrent users, with a latency of more than 2 seconds. This architecture includes Representational State Transfer Application Programming Interfaces (REST APIs) \cite{restful} endpoints for single and bulk inferences, as well as advanced workflows such as multi-step "tribunal" refinement. Experimental results confirm minimal overhead from container and scheduling activities and show that the approach scales reliably both for batch and interactive settings. We further illustrate real-world scenarios, including the deployment of chatbots with retrieval-augmented generation, which helps to demonstrate the flexibility and robustness of the architecture. The obtained results pave ways for significantly more efficient, responsive, and fault-tolerant LLM inference on large-scale HPC infrastructures.
\end{abstract}

\section{Introduction}

Large language models (LLMs) have transformed natural language processing by enhancing the understanding and generation of human-like text, which improves information retrieval and facilitates learning, as demonstrated in studies in \cite{hans1} and \cite{hans2}. However, deployment of such models poses some challenging computational challenges when such a system has to serve a significant number of users concurrently -- as the latest models in \cite{LLAMA32} and \cite{LLAMA31} already require high performance hardware to run individually. This paper provides an in-depth review of a SLURM-based scalable LLM HPC architecture by which such challenges are met through seamless integration of diverse LLM models within an HPC environment.

Our architecture leverages the powerful scheduling and resource management capabilities of SLURM to orchestrate inference job execution across heterogeneous HPC clusters. Dynamically generating SLURM batch scripts, it allocates required CPU, GPU, and memory resources for various inference engines such as Text Generation Inferences (TGIs) \cite{tgi} and virtual Large Language models (vLLMs) \cite{vllm} for their efficient parallel execution and optimal utilization of available hardware. It also allows for the integration of containerized microservices, using RESTful endpoints through FastAPI, to enable scalability and flexibility during client interactions that bridge the gap between hardware intensive HPC operations and user-facing applications. Web applications that include RESTful components (and service for large institutions) such as lecture significance evaluation in \cite{ICASSP2025} and LLM based chatbots need reliance of robust backends to run their workloads reliably \cite{lotsofpower}.

The paper is organized as follows: Section 2 surveys the related work in HPC scheduling and LLM inference frameworks. Section 3 describes the scalable engine architecture that includes HPC integration, network coordination, and load balancing. Section 4 focuses on its implementation: code integration, concurrency, and containerization. Section 5 presents experimental results on throughput, concurrency, overhead, and system behaviors across workloads. Section 6 concludes this work with a discussion on performance, robustness, scalability, and future research directions.



\section{Related works -- HPC and SLURM}


High Performance Computing (HPC) environments normally consist of large clusters of nodes that are interconnected by high-speed networks. These nodes consist of the conventional CPUs, different types of memory and specialized accelerators like GPUs. HPC systems are set up to maximize throughput, reducing the execution time for tasks that demand intensive hardware use and efficiently manage the available resources among numerous jobs.

The concept of job scheduling is an essential aspect of the Simple Linux Utility for Resource Management (SLURM) \cite{slurm} operation as it is responsible for allocating the resources available (nodes, GPUs, CPUs, RAM and storage) for each user request or query. Differently from conventional use, priority is based on the workload and queue time among the users.

Inference engines such as a TGI \cite{tgi} or vLLM \cite{vllm} and the required environment variables for the job. Once the job starts, SLURM spawns the requested jobs using the required hardware resources to finish the job. This can also include opening IP and port information for a Representational State Transfer Application Programming Interface (REST API) \cite{restful} endpoint.

The sequentially complexities described in sequence are recurrent when implementing the scalable engine with the HPC and SLURM interface. The first one includes network coordination and controller settings for readiness, statuses and intermediate results.

Large LLMs need distribution strategies that encompass multiple GPUs on a single node or multiple ones where model weights might require distribution among many GPUs (model parallelism) or each node handling a portion of the inference batch (data parallelism). For these cases, the call on large enough models must ensure correct orchestration for merging results or distributing new requests. 


Fault tolerance and scaling out are also considered in our architecture. Node failures or canceled jobs in execution might disrupt the system so the structure must be ready to re-queue and move jobs gracefully.



\begin{figure}[h]
    \centering
    \includegraphics[width=0.5\linewidth]{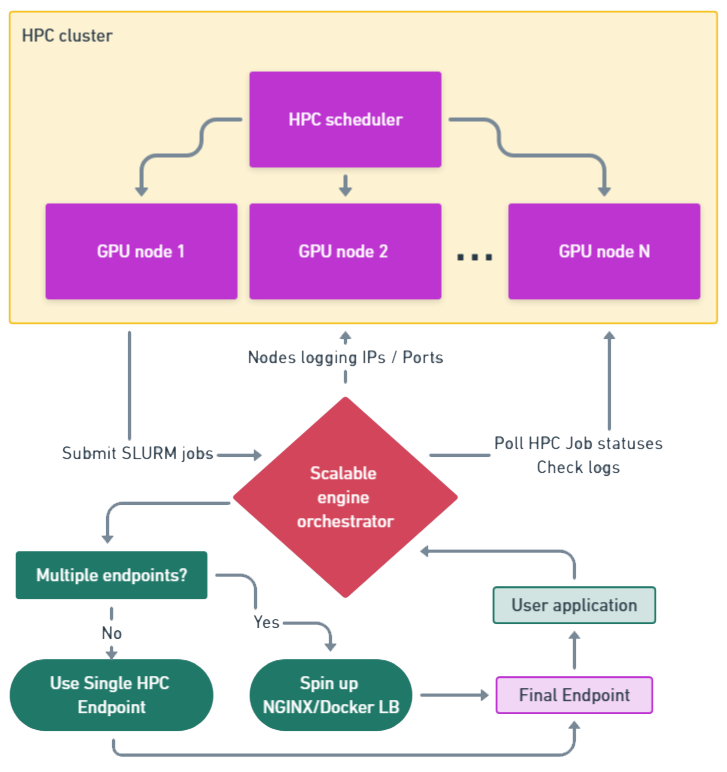}
    \caption{Diagram showcasing the activation path designed to provide the LLM endpoints.}
    \label{figureflow}
\end{figure}

While the inference frameworks innately handle low-level memory management and parallelism -- the current structure orchestrates the HPC interaction and concurrency logic according to Fig. \ref{figureflow}.

The user script or service can be any python code or microservice that initiates the scalable engine with a chosen model and required parameters to perform that job. The Scalable engine then reads the template and writes the parameters such as the inference engine, number of GPUs, model name and other hardware resources in the .slurm file. Afterwards, the conventional SLURM batch service is used to queue the job on the HPC system. The scheduler then handles the actual system queuing, resource allocation and dispatch of the jobs into the actual nodes \cite{slurm}. The HPC cluster would be granting exclusive or shared GPU nodes specified by the SLURM script according the required resources. Each node then runs a text generation server to handle these requests. The server logs the IPs and ports of such servers upon successful startup. The HPC job logs, titled "hosts file" register the endpoint addresses and allows the scalable engine to monitor which servers are up. The hosts file is parsed and retrieve the addresses for each HPC-based inference engine. If multiple endpoints for different services are found, such as a program using a REST API "server" to deploy another frontend server from gradio \cite{abid2019gradio} and streamlit \cite{streamlit2020streamlit}, the scalable engine programmatically creates an NGINX configuration, launching a container that unifies multiple endpoints into one load-balanced address. The final step returns one or more valid inference endpoints for all the created addresses to the user output. The script is then prepared to issue inference requests to any application either synchronously or asynchronously.

\section{Scalable Engine and HPC architecture}

The scalable engine uses the master-slave architecture to set up tasks in other nodes when according to the efficient hardware use algorithm set up by the load balancer from the scalable engine.

\begin{figure}[h]
    \centering
    \includegraphics[width=0.6\linewidth]{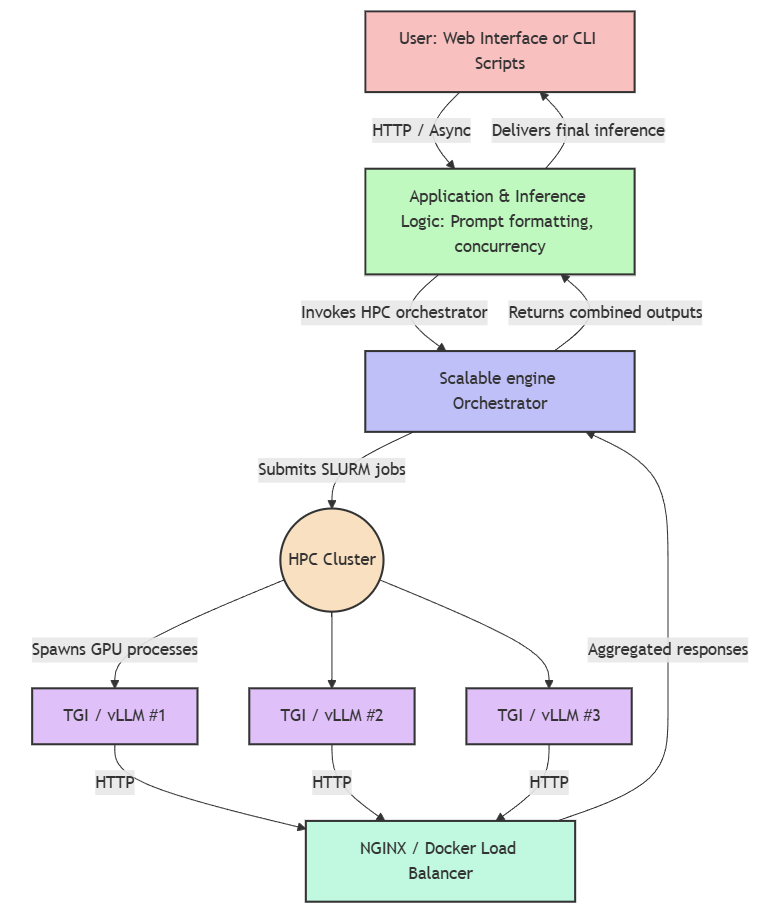} 
    \caption{Information flow for REST API call within the scalable engine.}
    \label{interface}
\end{figure}

Figure \ref{interface} shows the overall interface the scalable engine environment has with the HPC cluster. It represents how the system reacts when called upon to execute a REST API call.

The first input could be a FastAPI client -- \cite{fastapi}, Streamit or gradio interface or a custom chatbot making requests to the REST API address.
Table \ref{minres} represents the resources for the tested models.


\begin{table}[H]
\centering
\resizebox{\textwidth}{!}{%
\begin{tabular}{|l|c|c|c|c|}
\hline
\textbf{Model}        & \textbf{Parameters} & \textbf{CPU Cores} & \textbf{Memory (RAM)} & \textbf{GPU Requirements}             \\
\hline
Llama 3.1 8B \cite{LLAMA31}          & 8 billion            & 8 cores             & 16 GB                  & 1 GPU with 16 GB VRAM                  \\
\hline
Llama 3.1 70B \cite{LLAMA31}         & 70 billion           & 16 cores            & 128 GB                 & 2 GPUs with 80 GB VRAM each            \\
\hline
Llama 3.2 1B \cite{LLAMA32}         & 1 billion            & 4 cores             & 8 GB                   & Integrated GPU or 1 GPU with 2 GB VRAM  \\
\hline
Llama 3.2 3B \cite{LLAMA32}          & 3 billion            & 8 cores             & 16 GB                  & 1 GPU with 6 GB VRAM                    \\
\hline
\end{tabular}%
}
\caption{Minimum Hardware Requirements for Selected Llama Models}
\label{minres}
\end{table}

Jobs are subsequently submitted to the HPC cluster via the scalable engine orchestrator. The processes discovers the IP and ports of each started inference engine - and if multiple exist, set up a docker-based NGINX load balancer. The HPC cluster would then be managing the GPU resources by receiving the job submission from SLURM. Afterwards, an $N$ number of TGI/LLM/vLLM processes would run on the same or distinct nodes depending on the current workload. NGINX and the load balancer would distribute the inference calls across multiple HPC endpoints, receiving partial or complete outputs. The flow would aggregate the responses and send it back to the orchestrator, that would deliver to the application layer (already on the client side) and the output data from the running model. The application would display the outputted data in its desired form.


\section{Implementation, functionality and advanced use-cases}


In order to enable our system to operate seamlessly with RESTful microservices that mainly leverage the inference engine using web frontends, command-line tools or pipelines, a FastAPI \cite{fastapi} interface was developed to provide an asynchronous web layer on top of the HPC-based LLM inference. This is vital for any application that requires an easily deployable REST API to be deployed for a considerable number of users. Since machine learning pipelines can post multiple tasks via the batch endpoint, large scale text generation for functions like summarization, can be served in a single call.

For the developed architecture, the endpoint doesn't necessarily accept a single text input and triggers a single LLM inference, but the endpoint receives the call and runs a batch of predefined tasks in parallel, returning all completions together. Each API call uses python's asynchronous functionality to submit inference jobs concurrently to the HPC-based LLM. When the batch inference is invoked, the script launches multiple prompts simultaneously, yielding speedups proportional to the number of HPC workers.

A "tribunal" system ensures chatbot response quality by running a three-step HPC-LLM workflow (generate, critique, revise) guided by customizable "laws" that enforce standards like informal language or logical rigor. When system capacity allows, these laws trigger internal LLM debate to iteratively refine outputs, avoiding hallucinations and misalignment. To handle large inputs, prompts are split into N asynchronous chunks, processed in parallel by LLM instances, with summaries fed back to the tribunal layer for final review. All steps run concurrently to maintain low latency, and finalized responses are logged as accepted/rejected. During peak usage, the system bypasses advanced workflows, relying solely on the base model’s token limits. This balances quality control with scalability, adapting dynamically to user demand while preventing token overflow and minimizing manual tuning.
\section{Experiments and results -- Scalable engine saturation and use cases}

\begin{figure}[h]
  \centering
  \includegraphics[width=0.7\textwidth]{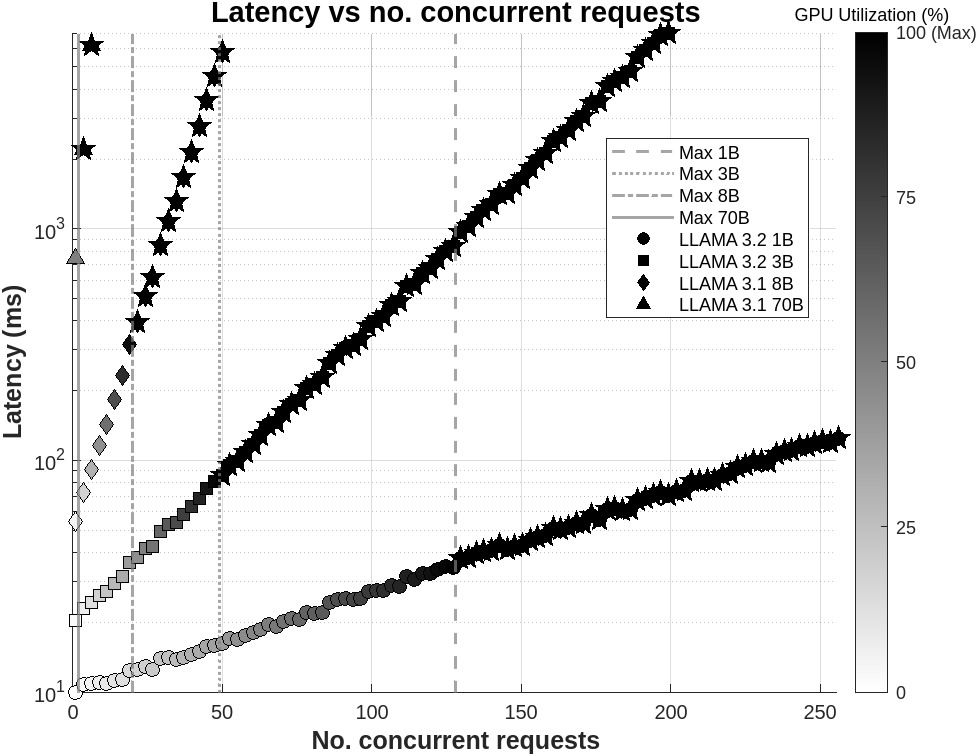}
  \caption{Representation of the performance obtained for the series of tests executed concurrently in order to test the limits of the current scalable engine. The vertical dotted lines are highlighted in gray are for the number of concurrent users and inference times: 1B (128, 36 ms), 3B (49, 85 ms), 8B (20, 336 ms), 70B (2, 2131 ms). The star symbol represents statuses operating under saturation and executing serialization. Startup time for each model is not considered.}
  \label{Diagram2}
\end{figure}
\begin{figure}[H]
    \centering
    \includegraphics[width=0.8\linewidth]{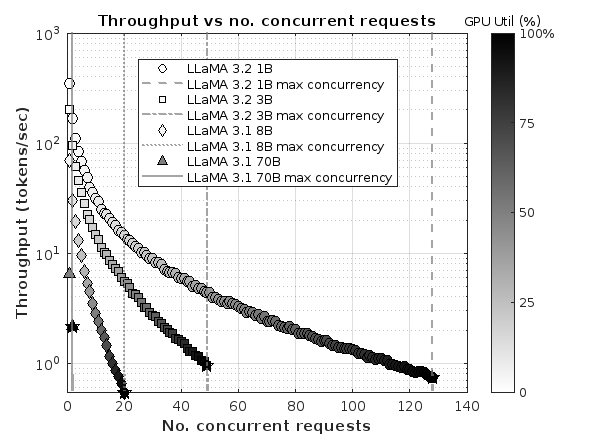}
    \caption{Representation of the token throughput per number of concurrent requests. The throughput is exposed up to the saturation point for each moment -- after which the behavior varies wildly as the throughput is influenced by the composite lag introduced by the queue-derived latency. Saturation is represented by a star when each individual model curve meets with the maximum concurrency line.}
    \label{throughput}
\end{figure}

Prompting the LLAMA 3.2 models (1B and 3B) \cite{LLAMA32} alongside the LLAMA 3.1 models (8B and 70B) \cite{LLAMA31} to translate the Lorem Ipsum \cite{lorem_ipsum} text from Latin to English, with 1024-token prompts in INT8 precision, provides insights into the performance of a scalable LLM inference system using a unified array of 4 NVIDIA A100 GPUs. 
Key metrics analyzed include the latency to generate the final token for each job, GPU load distribution, and the number of concurrent requests handled by the system. Figure \ref{Diagram2} highlights the system's behavior under maximum user load before task serialization begins. 
Inference initiation times vary significantly by model size, ranging from 10 ms for the 1B model to 750 ms for the 70B model. 
The saturation point, determined by the number of concurrent requests, decreases from 128 users for the 1B model to just 2 users for the 70B model. During this synthetic stress test, the jobs had the same priority and workload -- making the delivery of each output dependent on the First Input First Output (FIFO) policy. Figure \ref{throughput} represents how the throughput varies depending on the number of concurrent request the scalable engine is receiving simultaneously.

The throughput decreases as the number of concurrent requests increases since there is additional computational effort to deliver and orchestrate the requests produced by the GPUs -- making use of the CPUs and additional hardware resources -- for delivery for multiple clients. Similarly to the latency behavior, before the saturation point, each additional request doesn't drastically alter the throughput before the saturation point. Behavior after the saturation point would composite the delay derived from the additional latency the user experiences when waiting in the queue for the execution of each job. The execution of each job would then depend on the FIFO policy if all the jobs have the same workload for a synthetic stress test.

\begin{figure}[h]
    \centering
    \includegraphics[width=\linewidth]{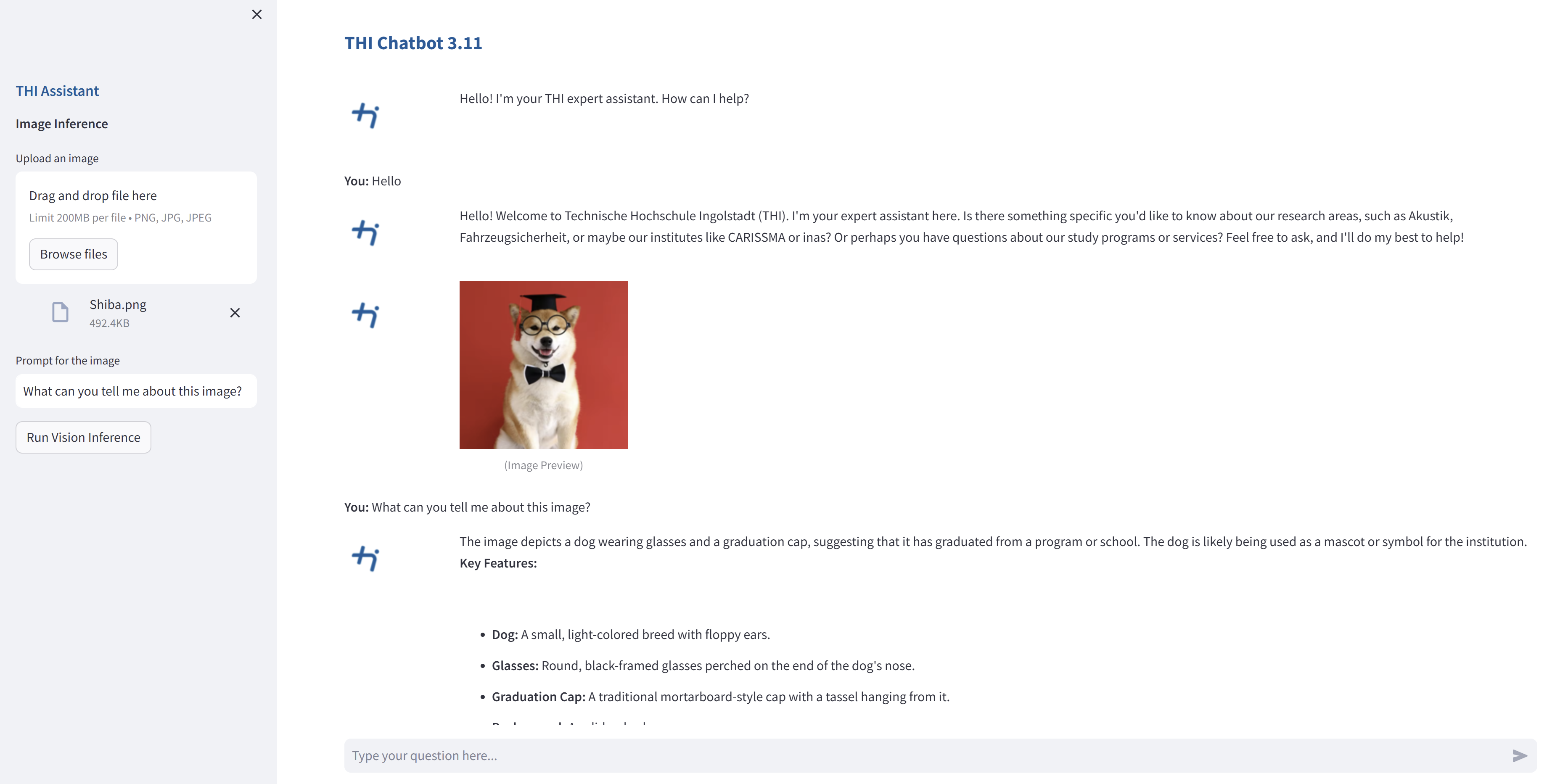}
    \caption{Representation of a chatbot window used for general purpose conversation with the integrated chatbot customized for Technische Hochschule Ingolstadt and vision inference.}
    \label{chatbot}
\end{figure}

Using the developed scalable engine, an example of a chatbot interface was developed for the user layer. It is focused on acquiring data usage in general university consultation settings. The chatbot scrapes through a limit of 100 subdomains from "thi.de" in order to build a chroma database stored locally in the client side for RAG \cite{lewis2020rag}. Figure \ref{chatbot} demonstrates the current deployment of the chatbot.

This use-case showcases how a client can develop their additional applications on top of the REST API by using resources that are served or built from their side - especially for customization or RAG tasks.

\section{Discussion}

Experiments show that SLURM-based HPC architecture effectively scales LLM inference but introduces trade-offs in model size and concurrency. Figure \ref{Diagram2} illustrates that smaller LLMs sustain higher concurrency and low latency, whereas larger models saturate with a considerable lower no. requests.

Figure \ref{throughput} confirms that within each model’s concurrency limit, the architecture efficiently exploits HPC resources for parallel inference, using containerized microservices, dynamic SLURM job submission, and load-balanced endpoints to maintain consistent performance. Overhead arises when job submission overlaps with queue wait times, but container spin-ups and scheduling remain manageable for both interactive and batch inferences.

A flexible RESTful interface and advanced “tribunal” workflow enable HPC-level resources to power use-cases like RAG-based question answering and iterative multi-step refinement. The chatbot scenario highlights how the framework supports additional client-side logic (e.g., summarization, chunking, and critique loops) while leveraging large-scale GPU clusters.

This unified orchestration for HPC-oriented LLM inference can be enhanced by smarter scheduling, advanced caching, fine-grained resource partitioning, and integration of new hardware and functions like in \cite{anderson}. Further efforts will focus on extending auto-parallelization for extreme-scale models and minimizing overhead during high concurrency across heterogeneous HPC environments.


\bibliographystyle{essv}
\bibliography{main.bib}

\end{document}